# Controlling End User Computing Applications
# - a case study

Jamie Chambers and John Hamill
34 Castlepark Road, Sandycove, Co Dublin, Ireland
chambersjamie@gmail.com

**ABSTRACT**

*We report the results of a project to control the use of end user computing tools for business critical applications in a banking environment. Several workstreams were employed in order to bring about a cultural change within the bank towards the use of spreadsheets and other end-user tools, covering policy development, awareness and skills training, inventory monitoring, user licensing, key risk metrics and mitigation approaches. The outcomes of these activities are discussed, and conclusions are drawn as to the need for appropriate organisational models to guide the use of these tools.*

## 1.  INTRODUCTION

The purpose of this paper is to share our experiences of a project which attempted to address the problems associated with the use of end user computing tools in a banking environment.

There was little published work to guide us. PricewaterhouseCoopers [2005] and Microsoft [2006] are useful on the processes to apply to spreadsheets, but do not look at the whole organisation. We believe that end user computing risk is first and foremost an example of operational risk so took our approach from that discipline.

Also, the scope of our project needed to be wider than the area of spreadsheet use. End users – that is to say, business users lacking professional IT training – now have many sophisticated computing tools and much computing power to deploy. So reporting programs, spreadsheets, databases and programming languages were all in scope and are generally referred to in this paper as End User Computing Applications (EUCAs). Given that the bulk of our critical EUCAs were spreadsheets, however, over the course of the project we concentrated primarily on this area of risk.

In addition, we felt that focussing solely on risk was only looking at half the picture. Our experience, reinforced by presentations at previous EuSpRIG conferences, has been that use of these tools is depressingly inefficient. We therefore also wanted to address the productivity aspects of end user computing, with the hope that those not overly concerned by the risk arguments might be at least be interested in potential benefits.

Our starting point was to show that we recognised both the risks and the productivity benefits that these tools bring. EUCAs are now a fundamental and useful part of the business environment, and there is no sense in attempting to eliminate them completely. By the same token, we had to recognise that the pervasive nature of end user computing meant that controlling these risks was likely to require considerable cultural change. Much of the project was therefore aimed at bringing about this change, no easy task when appreciation of the risks in these tools was not widespread.

## 2.  ORIGINS OF THE PROJECT

The Bank is a mid-sized international bank which had experienced rapid growth in its balance sheet and its use of structured instruments. While core systems were robust there was a constant need for systems to catch up with innovation and growth, which, combined with the general inclination amongst staff to use EUCAs, led to their proliferation.

An external audit comment was the primary stimulus for the project: the auditors remarked that there was a high level of dependency on complex spreadsheets particularly in the production of financial accounts. While the spreadsheet which was the primary focus of the comment was replaced, the general issue remained.

As the audit point touched several departments there was a need for central coordination to ensure a consistent organisation-wide approach. The Operational Risk department began to gather information about the use of EUCAs around the bank and the IT department started a dedicated project to address all the issues involved in end user computing activity.

We felt it was important to involve senior management to guide the project, approve the work and make recommendations to the executive, so we formed a Steering Committee, made up of heads of departments, with representation from both Front and Back Offices, and Internal Audit. This met monthly, and proved a useful source of ideas and support.

Much was achieved, as described below, but unfortunately during the course of the project there were some far-reaching executive changes, which led to a withdrawal of support for any centrally coordinated efforts to manage EUCA risk, and ultimately the suspension of the project.

**3. PROJECT WORKSTREAMS**

In aiming to change the way the Bank made use of end user tools we felt that both top-down and bottom-up approaches would be necessary, and identified a number of workstreams which would together bring about the necessary changes:

- Development of EUCA Policy & Control Standards
- Awareness Training
- Inventory of Critical EUCAs
- Risk Mitigation
- Licensing
- Skills Training
- Tools
- Deployment
- KRI Metrics

These activities and their outcomes are described below.

**3.1. Policy & Process**

The first activity for the project was to get agreement on the bank's attitude to EUCA risk, and to define the responsibilities and processes surrounding the use of EUCAs. This was set out in a policy, with the aim of:

- Placing responsibility for the risks arising from the use of EUCAs with managers.
- Understanding and reducing these risks through inventory and mitigation processes.

Though there are policy templates available, none of them was completely satisfactory. Their focus on Sarbanes-Oxley requirements was useful but too narrow for our needs, so we developed our own, and this was approved by the Steering Committee. We had to be careful to get managers to focus only on important EUCAs, so a central part of the policy was the development of a definition of business criticality.

Attempts have been made to classify spreadsheets and so assign levels of risk. For example PricewaterhouseCoopers [2004] have suggested rating spreadsheets on the basis of complexity (Low/Medium/High) and use (Operational, Analytical, Financial). A financial, high complexity spreadsheet would then be the most risky.

Unfortunately, even simple spreadsheets can cause large losses in an environment where very large transactions (> €1Bn) are commonplace, so an operational risk approach seemed to us to be more useful. We asked managers to think about impact and likelihood of loss through errors by assessing business criticality.

The criteria for an EUCA to be considered business critical and therefore to be included in the inventory were that it had to:

- Be used in financial or regulatory reporting functions, or
- Have the potential to impact financial statements by a material amount, or
- Be used to make decisions on significant investments or expenditures, or
- Be used to operate a core business function, or
- Support decisions in actual or potential revenue generation, or
- Be transmitted externally.

Having called, through the policy, for controls to be applied to critical applications we then defined what this meant through a Control Standards document. This laid out the minimum requirements for controls under the following headings:

- Version control
- Change control
- Access control
- Business Recovery
- Documentation
- Testing

The responsibility for implementing these standards was assigned to the business managers.

Finally, the Policy was owned by Operational Risk, while the Control Standards document was owned by IT. Enforcement lay with Internal Audit. All of these groups were heavily involved in the development of both policy and standards.

### 3.2. Inventory

The policy called for an inventory of business critical EUCAs to be maintained by managers. In order to get a grip on the use of EUCAs and to understand what was out there, all managers were sent a spreadsheet template and requested to list their critical applications. This would demonstrate recognition of their critical applications, and help to monitor their progress in controlling them or migrating functionality into core IT systems. In addition to basic information for each EUCA (name, owner, end user tool employed), we also asked for information which could help us understand risk (extent of any QA process, modification frequency, number of users) as well as the type of use and level of complexity. Responses were compiled into one Bank-wide inventory.

Inevitably this was a very difficult process: we had to put in a lot of effort to get managers to comply, and even then were not sure that all important applications had been acknowledged. This was not necessarily sinister: sometimes end user applications were so embedded in working practices that they had ceased to be recognised as such, or were believed to be the responsibility of others. We discovered several instances of highly critical applications which didn't appear on anyone's inventory as no one felt they owned them.

Problems in assessing business-criticality also lead to either over- or under-declaration of applications. As is common with operational risk self-assessments, managers often needed guidance on how to use the criticality guidelines. Finally, cross checking of inventories was needed to remove duplication and ensure clear ownership.

Maintaining a manual inventory of EUCAs is a formidable task. The difficulties in devising automated means of EUCA risk measurement make it unlikely that discovery tools will completely replace the need for some form of self-assessment, but we see them playing an important role in informing the inventory process. Without an automated solution, the maintenance of the inventory requires compliance from managers backed up by the threat of internal audit sanctions.

In addition to the production of an inventory of existing applications we also included in the policy a mechanism for the approval of new ones which required the departmental manager to consult with the Head of IT. This was to ensure that no redundant EUCAs were created and those that were created were immediately included in the inventory.

Finally, though the main function of the EUCA inventory is to understand the population of business critical applications, it is also provides a unique view on the weaknesses in central IT systems. We therefore shared information from the inventory with those responsible for strategic IT reviews.

### 3.3. Awareness Training

Having established the policy, we held dedicated awareness training sessions for all managers to explain their obligations under it. The sessions also gave us the opportunity to educate managers on the nature of EUCA risk, so we provided them with:

- evidence of spreadsheet risk events (courtesy of the EuSpRIG website)
- an overview of uses of spreadsheets [Grossman, Mehrotra & Özlük, 2007]
- evidence of spreadsheet problems [Panko, 1998]
- data on spreadsheets in the context of human error research [Panko, 2007]

These sessions were well attended and achieved the goal of making managers think about the problem in ways most of them had not previously considered.

In fact managers were not completely unaware of EUCA risk, as shown by their often slightly defensive attitude in discussions over their use of end user tools. The heavy users of spreadsheets were somewhat guilty over their dependency, but felt they lacked alternatives. They certainly lacked any awareness of the means of managing the risks arising from their spreadsheets, and voiced concerns that they would be unable to implement the policy without considerable help.

### 3.4. Mitigation Activities

The EUCA policy required managers to apply the minimum control standards to their business critical EUCAs or else to migrate that functionality into core IT systems. We therefore began to work with managers to develop programmes for them to accomplish this.

In most cases, migration to core systems was not feasible in the short term, given the lengthy queue for IT development work. This was, of course, no surprise: EUCAs are often developed to fill the gap between IT system capabilities and current business requirements.

Of more concern was the finding that EUCA development often proceeded without checking whether the required functionality already existed in the central Bank systems. We found cases where considerable manual effort was put into maintaining data and producing reports which could have easily been produced from core systems with the added advantage of benefiting from the quality control applied by IT.

Applying controls was the only short to medium term option for most EUCAs, and here we ran into two sorts of problems. First, managers needed considerable assistance in implementing controls as they lacked the requisite skills. To those with experience of software development processes, the control standards made no unusual demands, but these skills are not widely available and we underestimated the effort needed to educate users to apply them. At the time the project was suspended we were working to encourage managers to invest in the skills needed to implement the standards within their teams.

Second, the process of producing a particular application was often so manual that applying the standards would not have been feasible. An example was a reconciliation spreadsheet which compared the trade records within the main trade processing system with those in the general ledger system. The spreadsheet had to load reports from the two systems, parse them, and construct pivot tables to make the comparison. This was automated to some degree, with a 400-line (recorded!) VB macro. But once the macro had run, heavy manual editing was needed to make good the deficiencies in the original reports. To put a set of controls around the process as it existed would have made no sense: in effect, the only way to test the process would have been to produce a duplicate report independently. Instead the application had to be reconstructed in order to be able to demonstrate control at each stage of the editing process.

Our conclusion is that it will often not be possible to simply wrap controls around an existing application; mitigating the risks of EUCAs may mean reworking the processes used to generate them. Since the skills required to do this rarely exist in end user departments, mitigating EUCA risks will often require considerable assistance from IT.

In the latter part of the project we worked closely with the Finance Department to apply controls to several key spreadsheets. We found that combining the productivity and risk approaches garnered most cooperation. Some significant gains were made in both risk control and productivity and some knowledge of good practice was passed on to end users. The lesson learned was that a very hands-on approach is needed to spread the message and create departmental centres of expertise.

**3.5.     Skills**

It is now widely recognised that users of EUCA tools are generally under-trained (not to mention untrained). Our organisation, we believe, is not atypical in providing any member of staff with unconstrained use of a complete Microsoft Office suite of tools, as a matter of course. We felt that a bottom-up approach to controlling EUCA risks might be useful, so organised a number of sessions of 'Excel Good Practice Training' with an open invitation to all staff.

In a busy environment it is hard to get much of people's time, so we restricted the sessions to 2 hours, aiming to provoke staff to think about their use of Excel, both from risk and productivity perspectives, and to give them some tools and methods to use in their daily work.

Deciding on the content of the sessions was of course difficult, but eventually we settled on the following agenda, to get users to:

- handle large datasets safely and efficiently

- use formulas and functions more effectively
- incorporate error checking in their work
- analyse and design their applications
- make better use of security features, and be aware of their limitations

We took the opportunity, as with the Awareness sessions, to set the problem context, and show users what would be required of them by the new Control Standards. We reviewed participants' training and experience at the outset, and provided them with resources such as example spreadsheets, on the corporate intranet.

The three sessions were very well attended. Over 70 people took part, and many more expressed interest but were unable to come. All attendees made regular use of Excel, but only 5 had ever been on an Excel training course. The experience review was also surprising. Many used formulas, but rarely anything more complex than IF. Very few made use of the Data Validation or Conditional Formatting functions, only two people knew about the Go To Special function, and no one knew of the Formula Auditing function, all of which should be key tools in error checking processes.

A two hour training session can achieve little in the way of behaviour modification. The most that we could hope for is that our sessions would raise awareness amongst users of the value for them in increasing their knowledge of Excel.

It was clear that there was a tremendous appetite amongst staff to learn more about Excel, and the Bank had a generous training budget. Why then the lack of skills? The Bank is not alone in providing users with a range of tools but not the formal training to use them effectively. The Excel training that had been done had focussed on skills rather than risk and control. Our conclusion is that an organisational view on appropriate skill levels is needed, accompanied by a dedicated training programme for all users. We believe that this could be justified from both risk reduction and productivity improvement perspectives, but it is difficult to make this argument in the absence of any quantitative cost-benefit analysis.

We also constituted an Excel User Group comprised of six or so more experienced users of Excel to act as an internal source of expertise and a discussion forum for problems. This also helped to raise the profile of the project, and for the first time gave users a focal point within the Bank to direct their questions and concerns via email.

### 3.6. Tools

At the beginning of the project we expected that spreadsheet control tools would play an important part in our risk reduction strategies and we put some effort into researching the alternatives. We hoped to recommend tools at both user-level (for example, Spreadsheet Professional or EXChecker) and entity-level (such as Cluster 7 or Compassoft). There was no use, and little knowledge, of any of these tools in the Bank when we began.

Unfortunately this workstream made little progress, once it became clear that management enthusiasm for centrally-managed EUCA controls was lacking. Nevertheless we drew some conclusions from our work.

It is clear that entity-level tools have the capability to institute controls, such as change control and versioning, over large numbers of spreadsheets for very little effort, and can obviate much of the work required to institute systems of control and train users to operate them. They can also make useful contributions to the Inventory process through discovery of spreadsheet populations, though they can only indirectly assist with assessments of business-criticality.

User-level tools can contribute significantly to improving the quality of individual spreadsheets and to enforcing corporate standards.

However, both kinds of tool require central functions to take responsibility for operation, management, administration, the development of standards and provision of training. Though the IT function in our organisation would have accepted the operational responsibilities, it was not clear who would take overall control, and nor was it clear who would take budgetary responsibility.

### 3.7. Deployment

At present we (and many other organisations in our experience) deliver the full functionality of end user tools to users without ascertaining their needs or competencies, and we do not attempt to configure the default settings.

Should all users get all tools automatically? At the very least we believe this should be an explicit decision, based on risk as well as other factors. Granting users open access has benefits - they may use the tools effectively - but undoubtedly leads to the proliferation of uncontrolled applications. Some evidence of the trade-off between convenience and risk-reduction would be useful here, but unfortunately we cannot provide it: initial discussions with the IT Infrastructure department showed that they were keen neither on shouldering the administrative burden, nor on becoming a central control function in this manner, and we made no further progress.

In fact our proposals were not onerous: at the least there should be a policy as regards the provision of EUC tools, and that advanced tools (we thought for example of Access, or the VB editor in Excel) should be made available only on request via managers, but without a highly bureaucratic approval mechanism. In this way managers would become aware of the activities of their staff in using complex end user tools.

We would also have been able to promote corporate standards in spreadsheet construction through customised templates, and to ensure appropriate default settings.

### 3.8. Licensing

The logic of the need to license or certify those that operate critical EUCAs seems to us to be unassailable: if the risks of operating spreadsheets are assessed to be significant, and if the managers who own those spreadsheets are to be held to account for any losses arising from them, then they will want to know that the operators are properly qualified. This will mean certification of some sort: passing an external examination to demonstrate achievement of the required standard.

However, as others have noted, there is resistance to this idea, as there is to any move to promote standards in a laissez-faire environment. Our steering committee agreed that bus-drivers should be licensed to a higher level of skill given the higher risks of their work compared with private cars, yet were uncomfortable with our proposal for introducing spreadsheet certification, evidence yet again of the underestimation of spreadsheet risk.

Our way of promoting standards in the short-term, and gaining acceptance to the idea, was to offer voluntary certification to users. This was fine in theory, but actually there is very little on offer in this area. Eventually we made contact with Q-Validus who were developing the kind of certification we were interested in, designed to help users assure good spreadsheet design, usage and control ('Spreadsheet Safe').

We ran a pilot course and exam with seven of our users, of varying experience, from audit, and the front, middle and back offices. Following up two months later, all found the course useful and to have made a difference to the way they approached their spreadsheet work.

### 3.9. KRI Metrics

Key Risk Indicators (KRI) are measurements of factors drawn from operational environments whose magnitude or direction of movement over time point to changes in operational risks. The Operational Risk department delivers a Key Risk Indicator reporting pack each month to senior management, with contributions from all departments. We felt that developing some metrics around the use of EUCAs would be a useful way to indicate the level and, over time, the direction of this risk and hence the success (or otherwise) of our workstreams.

Devising useful metrics is not always easy: the total number of critical EUCAs obviously has some influence on risk and was the first output from the early inventory. But this is an indicator of the absolute level of risk, not the controlled level. Therefore we gathered more information in the second round of the inventory on the level of controls applied to the critical applications and so could report, by department, the proportion that had no QA process.

This gave us a way to illustrate to management how seriously the risk is being taken by managers. Ideally we would like information on actual EUCA risk events – instances of losses, either direct or indirect, caused by errors, to give a more immediate assessment of risk. This may be an area where an entity-level tool could provide useful statistics, for example on unsafe practices. Automated collection of the indicator data is also highly desirable if metrics are to be kept up-to-date.

### 4. SUMMARY AND CONCLUSIONS

As expected, we found that EUCA risk was poorly understood, and rarely controlled in any way around the Bank. Our observations echoed those of Croll [Croll, 2005]: 'there is almost no spreadsheet software quality assurance or appreciation of the software development life cycle as it might relate to spreadsheets'. Though we encountered general acknowledgement that the uncontrolled use of EUCAs was a source of operational risk, it was interesting to note that few managers felt responsibility, believing their applications to be well controlled, or unimportant.

No attempt was made to ensure staff were qualified in the development of EUCAs to a level commensurate with their responsibilities. Managers were grateful when their staff constructed applications to address processing and reporting issues, but had no framework for supporting, controlling, managing or even promoting these activities.

As EUCAs become more powerful and more widely available, and as areas which were once the province of programmers alone become accessible to untrained general users, these problems can only grow in size.

Despite its early termination, our project made significant progress in controlling the use of end user tools throughout the Bank. It established a framework (Policy and Standards) to control the use and proliferation of EUCAs, raised awareness of EUCA and particularly spreadsheet risk among managers, and gave them techniques for assessing that risk in their own departments. It identified an appetite for both skills and risk training which will, we hope, lead to its provision by individual managers. Combining productivity training with risk awareness training seemed to produce the best response: our users accepted that they needed to reduce the risks in their applications but were eager to use their end user tools more efficiently as this brought them immediate benefits. For the first time an inventory of critical EUCAs across the organisation was created and this in turn allowed an initial assessment of the risk facing the organisation which had not previously been possible.

However, while the project dealt with the open audit point and made some progress on a general approach to the EUCA issue, the problem of ownership (and hence of budgeting) meant that the project ended prematurely. This key issue will have to be addressed in any similar project elsewhere.

Primary ownership of the end user computing problem was agreed to belong to the individual managers who ran the risk. However this left open the question of how the problem should be centrally tackled, as it has to be. Departmental initiatives, while useful, cannot fully address the organisation-wide aspects of the problem which were evident in all of our workstreams. These include maintenance of an inventory of EUCAs, training and licensing of users, development of policy and standards, controlled roll out of tools and overall coordination. Candidates for a central coordinating role in our case were Operational Risk and IT or a combination of both, but neither in the end was willing to take this responsibility. A standardised approach to the problem, dividing the responsibilities between IT, Operational Risk and departmental managers, promoted by groups such as EuSpRIG, could help individual organisations both to recognise and to tackle the risk in a coherent way.

Ultimately we formed the view that the central difficulty in controlling end user applications is not discovering them or mitigating their risks, challenging though these are. The central difficulty lies in formulating, and getting accepted, an organisational response - a standard model for the use of end user computing applications - that addresses all the issues raised by the use of end user tools, including those of productivity. All aspects of our project required central coordination, but there is as yet no consensus as to the appropriate location of that coordination function [Lepus, 2006]. Despite all the evidence of risk and inefficiency, few managers set about addressing these problems, largely, we believe, because they lack the skills and support to do so.

In our view, without such an organisational model effective and safe use of end user computing tools is unlikely to be achieved. Put in a wider context, the conclusion we draw is that the development of this technology has outstripped the evolution of work practices to cope with it, as has been observed in previous periods of technological innovation [David, 1990]. We will neither control the risks nor realise the productive potential of end user computing tools on a sustained basis across organisations until such a model has been adopted.